\documentclass[sigconf,nonacm]{acmart}

\usepackage{amsmath,amsthm,mathtools}
\usepackage{booktabs}
\usepackage{array}
\usepackage{microtype}
\usepackage{xcolor}
\usepackage{enumitem}
\usepackage{graphicx}
\usepackage{algorithm}
\usepackage{url}
\usepackage{xspace}
\usepackage{placeins}

\newtheorem{theorem}{Theorem}
\newtheorem{proposition}{Proposition}
\newtheorem{corollary}{Corollary}
\newtheorem{lemma}{Lemma}
\newcommand{\detv}[2]{\det(#1,#2)}
\newcommand{\J}{J}
\newcommand{\LL}{\mathcal{L}}
\newcommand{\R}{\mathbb{R}}
\newcommand{\norm}[1]{\lVert #1\rVert}
\newcommand{\CloseSort}{\mbox{\textsc{CloseSort}}\xspace}
\setcopyright{none}
\title[Close and Sort]{Close and Sort: Exact Maximum-Likelihood Aggregation of Cyclic Rank-Two Preferences}

\newcommand{\aff}[1]{\texorpdfstring{\textsuperscript{#1}}{}}
\author{Kehan Liu\aff{1}\texorpdfstring{\qquad}{, }Zhaoyun Chen\aff{1}\texorpdfstring{\qquad}{, }Xianquan Yan\aff{1,2,}}
\authornote{Corresponding author: \texttt{yanx@u.nus.edu}}
\affiliation{%
  \institution{Department of \{\aff{1}Physics, \aff{2}Computer Science\}}
  \city{National University of Singapore}
  \country{Singapore}}

\begin{document}

\begin{abstract}
Pairwise probabilities arise in rank aggregation, recommendation, and human or
LLM evaluation, yet converting them into a maximum-likelihood list is a
weighted Kemeny problem and is NP-hard in general.  Known exact algorithms for
rank-two tournaments use only comparison signs and do not optimize this
likelihood.  We show that the weighted problem is exactly solvable when the
skew-symmetric log-odds matrix $W_{ij}=\log(p_{ij}/p_{ji})$ has rank two,
including when the induced preferences contain cycles.  Writing
$W_{ij}=\det(h_i,h_j)$ and adding the closing vector $h_0=-\sum_i h_i$ turns
the objective into twice the signed area of a closed polygon.  This geometry
yields \CloseSort: polar-sort the $n+1$ edge vectors and cut after
$h_0$.  From the factors, it returns a global maximum-likelihood order in
$O(n\log n)$ time.

The same geometry identifies the runner-up and gives a stability certificate
beyond exact rank two.  Let $\delta$ be the smallest determinant of two
consecutive edges in the closed angular cycle.  When $\delta>0$, it is exactly
the best-to-runner-up log-likelihood gap.  For
$W=W_0+E$ with an arbitrary skew residual, the \CloseSort order of
$W_0$ has regret at most $\sum_{i<j}|E_{ij}|$ on $W$ and remains globally
optimal whenever this quantity is below $\delta$; this threshold is worst-case
tight.

Exact optimization on synthetic instances matches the predicted
optimum, runner-up gap, and tight stability threshold.  In simulations of
5,120 determinant-logit profiles, increasing the number of observations per
pair from 4 to 64 raises the fraction on which rank-two projection outperforms
the empirical plug-in optimum from 68.6\% to 85.7\%, with both rankings
evaluated against the latent population log-odds.  On 3,600 Arena-Hard
LLM-judge profiles of seven systems and five questions each, drawn from 63
systems and ten judges, the leading rank-two component captures 89.2\% of the
log-odds energy.  Among profiles containing a
directed majority triangle, \CloseSort reduces normalized regret by 44\%
relative to Borda and achieves a positive mean normalized objective gain over
each of eight baselines.  For 18-item rankings with
two-dimensional factors supplied, its angular-sort inference is 8,407 times
faster than exact dynamic programming.

Code is available at
\href{https://github.com/sarinstein-yan/close-and-sort}{\textcolor{ACMDarkBlue}{github.com/sarinstein-yan/close-and-sort}}.
\end{abstract}


\maketitle
{\let\thefootnote\relax\footnotetext{Accepted at the 20th ACM International Conference on Web Search and Data Mining (WSDM~'27), February 15--19, 2027, Hong Kong, China.}}

\section{Introduction}

A search or recommendation system often has pairwise evidence---clicks, ranker
disagreements, human preference votes, or LLM-as-a-judge comparisons---but must
display one list.  Rank aggregation is a core primitive in group recommendation
and metasearch~\cite{baltrunas2010group,dwork2001rank}, and the same need now
arises in LLM evaluation.  LLMs are widely used for automatic NLG
evaluation~\cite{liu2023geval,bavaresco2025llms}.  Pairwise comparative
assessment is a practical LLM evaluation protocol~\cite{liusie2024comparative}
and underlies MT-Bench and Chatbot Arena~\cite{zheng2023judging}, and
system-level benchmarks compare dozens of generators under multiple judges and
scoring protocols~\cite{gera2025justrank}.  These comparisons need not be
transitive: recent work documents non-transitive LLM-judge preferences and
their effects on induced rankings~\cite{xu2025nontransitivity}, so a single
list may have to summarize cyclic evidence.

Let $p_{ij}$ be the probability that item $i$ should precede $j$, with
$p_{ij}+p_{ji}=1$.  Under independent pairwise observations, the likelihood of
an order $\pi$ is
\begin{equation}
  \LL(\pi)=\prod_{r<s}p_{\pi_r\pi_s}.
  \label{eq:likelihood}
\end{equation}
Maximizing~\eqref{eq:likelihood} is the weighted Kemeny or linear-ordering
problem~\cite{marti2011linear}.  The classical Kemeny rule admits a
maximum-likelihood interpretation under Condorcet-style
noise~\cite{young1988condorcet}, and agreement maximization over pairwise
preferences is a standard learning-to-rank
formulation~\cite{cohen1999learning}.  The problem is
NP-hard~\cite{bartholdi1989voting,alon2006ranking,charbit2007minimum}; exact dynamic
programming costs $O(n2^n)$ and general weighted tournaments have approximation
schemes rather than a simple exact
solver~\cite{fischer2016weighted,kenyon2007ptas}.

A natural tractable exception is known for rank-two tournaments.  A tournament
is rank two when its edge directions are the signs of entries of a rank-two
skew-symmetric matrix.  Prior work shows that minimum feedback arc set (MFAST)
can be solved exactly on this class, including tournaments with directed
cycles~\cite{veerathu2021structure,rajkumar2022theory}.  MFAST seeks an order
with the fewest backward arcs, so it uses only comparison directions and
assigns every reversed arc unit cost.

Maximum likelihood poses a different weighted problem.  Let
$W_{ij}=\log(p_{ij}/p_{ji})$ be the log-odds matrix.  Reversing a preferred pair
changes log-likelihood by $|W_{ij}|$, so comparison magnitudes affect the
optimum.  Indeed, the MFAST-optimal order need not maximize
\eqref{eq:likelihood} (Section~\ref{sec:separation}).

Our main result extends exact rank-two tractability from sign-only MFAST to
weighted maximum likelihood.  The extension follows from a planar
representation: every rank-two log-odds matrix $W=-W^\top$ admits factors
$h_i\in\R^2$ such that $W_{ij}=\det(h_i,h_j)$.  Adding the closing
vector $h_0=-\sum_i h_i$ then turns the weighted objective into twice the signed
area of a closed polygon with prescribed edge vectors $h_0,h_1,\ldots,h_n$.  A
classical edge-rearrangement result makes polar-angle order maximize this
area~\cite{bader2020horofunction}; cutting the resulting cycle after the
distinguished edge $h_0$ returns a linear item order.  This construction yields
\CloseSort, a global maximum-likelihood algorithm that runs in
$O(n\log n)$ time from the factors.

The class includes transitive and cyclic models.  Bradley--Terry is
the transitive special case: its log-odds satisfy $W_{ij}=s_i-s_j$, and
\CloseSort simply sorts the scores $s_i$~\cite{bradley1952rank}.
General factors can produce directed cycles.  Their determinant log-odds
coincide with the one-dimensional, bias-free blade-chest-inner model of Chen
and Joachims~\cite{chen2016intransitivity}.  \CloseSort therefore returns an
exact maximum-likelihood list for that model.

The polygon also identifies the runner-up and controls departures from exact
rank two.  Let $\delta$ be the smallest determinant of two consecutive edges in
the closed angular cycle.  When $\delta>0$, it is exactly the
best-to-runner-up log-likelihood gap.  For $W=W_0+E$, the order obtained from the
rank-two component $W_0$ has log-likelihood regret at most
$\sum_{i<j}|E_{ij}|$ on $W$ and remains globally optimal whenever this sum is
below $\delta$; this threshold is worst-case tight.

Empirical log-odds are estimated from finite comparisons and need not be
exactly rank two.  We therefore study rank-two structure both as a denoising
model when the population log-odds are rank two and as an approximation to
real matrices, where the resulting order is evaluated on the original
weighted objective.

Our contributions are:
\begin{enumerate}[leftmargin=*,itemsep=2pt]
\item \textbf{Exact weighted aggregation.} We reduce rank-two weighted Kemeny
  to maximum-area ordering of prescribed polygon edges.  The closing vector and
  its distinguished cut give a global maximum-likelihood order in
  $O(n\log n)$ time from the factors
  (Theorem~\ref{thm:exact}).
\item \textbf{Exact gap and residual stability.} The same polygon gives the
  exact runner-up gap when $\delta>0$, a regret bound for arbitrary skew
  residuals, and a worst-case-tight stability certificate
  (Theorem~\ref{thm:stability}).
\item \textbf{From exact factors to observed log-odds.} Exhaustive enumeration
  and exact MILP confirm the predicted optimum, runner-up gap, and stability
  threshold.  Under determinant-logit sampling, rank-two projection has lower
  population regret than the empirical plug-in optimum on 85.7\% of profiles
  at 64 observations per pair.  On Arena-Hard LLM-judge profiles containing a
  directed majority triangle, \CloseSort reduces normalized regret by 44\%
  relative to Borda and achieves a positive mean normalized objective gain over
  each of eight baselines.  Runtime experiments confirm the computational
  advantage of angular sorting over exact dynamic programming when the
  two-dimensional factors are given.
\end{enumerate}

Table~\ref{tab:landscape} compares the closest exact-ranking settings.  The
rank-two MFAST row uses the $O(n^2)$ Quicksort-and-cyclic-shift
method~\cite{rajkumar2022theory}.

\begin{table}[t]
\centering
\caption{Comparison of exact-ranking methods for pairwise preferences under
rank-two structure, alongside unrestricted general Kemeny.  BT sort and
\CloseSort start from scores or factors; rank-two MFAST starts from tournament
signs.}
\label{tab:landscape}
\small
\begin{tabular}{lccc}
\toprule
Method & Cycles & Exact MLE & Complexity\\
\midrule
BT sort & $\times$ & $\checkmark$ & $O(n\log n)$\\
Rank-two MFAST & $\checkmark$ & $\times$ & $O(n^2)$\\
\CloseSort & $\checkmark$ & $\checkmark$ & $O(n\log n)$\\
General Kemeny (unrestricted) & $\checkmark$ & $\checkmark$ & NP-hard\\
\bottomrule
\end{tabular}
\end{table}

\section{Setup and likelihood reduction}

There are $n$ items.  We assume $p_{ij}\in(0,1)$,
$p_{ij}+p_{ji}=1$, and define the log-odds
\begin{equation}
 W_{ij}=\log\frac{p_{ij}}{p_{ji}}, \qquad W=-W^\top.
 \label{eq:logodds}
\end{equation}
For a permutation $\pi$, define
\begin{equation}
 \J_W(\pi)=\sum_{r<s}W_{\pi_r\pi_s}.
 \label{eq:J}
\end{equation}

For any pair $(\pi_r,\pi_s)$ with $r<s$, decompose its
log-probability contribution as
\[
 \log p_{\pi_r\pi_s}
 =\tfrac12\bigl(\log p_{\pi_r\pi_s}+\log p_{\pi_s\pi_r}\bigr)
  +\tfrac12 W_{\pi_r\pi_s}.
\]
The symmetric term depends only on the unordered pair $\{i,j\}$, not on
which item leads.  Summing over all $\binom n2$ pairs, the symmetric
contributions form a constant $C(W)$ independent of $\pi$, and the
antisymmetric terms sum to $\tfrac12\J_W(\pi)$.  Thus
\begin{equation}
 \log\LL(\pi)=C(W)+\tfrac12\J_W(\pi).
 \label{eq:likelihood-reduction}
\end{equation}

Maximum likelihood and maximum $\J$ are therefore identical.  For a skew matrix of
rank two there are $h_i=(x_i,y_i)\in\R^2$ such that
\begin{equation}
 W_{ij}=\detv{h_i}{h_j}=x_i y_j-y_i x_j.
 \label{eq:detfactor}
\end{equation}
\begin{proposition}[Rank-two factors]\label{prop:factors}
For nonzero skew $W$ of rank two, choose $p,q$ with $W_{pq}\ne0$ and set
\begin{equation}
 h_i=\left(W_{ip},\frac{W_{iq}}{W_{pq}}\right).
 \label{eq:explicitfactor}
\end{equation}
Then~\eqref{eq:detfactor} holds.  Moreover, any two full-rank determinant
factorizations differ by an orientation-preserving linear transformation, so
they induce the same cyclic polar order.
\end{proposition}

\begin{proof}
The rank-two Pl\"ucker identity gives
\begin{equation*}
 W_{ip}W_{jq}-W_{iq}W_{jp}=W_{ij}W_{pq},
\end{equation*}
which proves the first statement.
For the second, write $W=H\Omega H^\top$ with rows $h_i$ and
$\Omega=\left(\begin{smallmatrix}0&1\\-1&0\end{smallmatrix}\right)$.  Two full-rank
factor matrices have the same column space $\operatorname{range}(W)$, so
$H'=HA$ for an invertible $2\times2$ matrix $A$.  Equality of the two
factorizations implies $A\Omega A^\top=\Omega$.  In two dimensions this is equivalent to
$\det(A)=1$.  Thus $A$ preserves orientation, determinants, and cyclic polar
order.
\end{proof}

Equation~\eqref{eq:explicitfactor} recovers factors from a dense matrix in
$O(n^2)$ time including rank verification.  When a determinant-logit model
(Section~\ref{sec:detlogit}) supplies the factors directly, constructing the
dense matrix is unnecessary.

\section{Close and sort: the exact algorithm}
\label{sec:exact}

\begin{figure}[t]
  \centering
  \includegraphics[width=\linewidth]{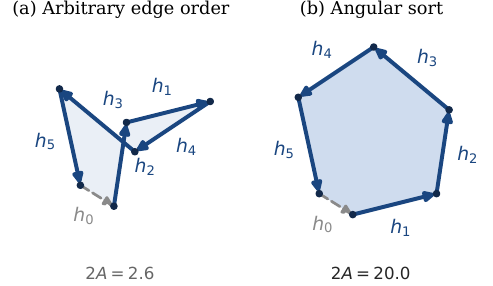}
  \caption{The augmented edge vectors sum to zero.
  (a)~An arbitrary order traces a self-intersecting polygon with small signed
  area.  (b)~Angular sorting produces a weakly convex, maximum-area polygon;
  removing the closing edge $h_0$ (dashed) yields the item ranking.
  The objective equals twice the signed area.}
  \Description{Two-dimensional comparison factors are closed into a polygon
  and arranged by polar angle to maximize its signed area.}
  \label{fig:geometry}
\end{figure}

The key observation is geometric.  Augmenting the $n$ factor vectors with
a closing vector $h_0=-\sum_{i=1}^n h_i$ produces $n+1$ vectors that sum to
zero.  Any ordering of these vectors traces a closed polygon, and $\J_W(\pi)$
equals twice its signed area.  Angular sorting of prescribed edge vectors
maximizes polygon area; a single cut after the distinguished edge $h_0$
converts the optimal cyclic order into a linear ranking.  We call this
procedure \CloseSort (Algorithm~\ref{alg:closesort});
Figure~\ref{fig:geometry} illustrates its polygon construction.

\begin{algorithm}[t]
\small
\hrule\vspace{3pt}
\textbf{\CloseSort$(h_1,\ldots,h_n)$}\\[-2pt]
\begin{enumerate}[leftmargin=18pt,itemsep=0pt,topsep=2pt]
\item Set $h_0\leftarrow-\sum_{i=1}^n h_i$.
\item Polar-sort $0,1,\ldots,n$ counterclockwise by half-plane and determinant
  comparisons.
\item Rotate the cyclic indices until $0$ is first.
\item Delete $0$ and return the remaining indices.
\end{enumerate}
\vspace{-2pt}\hrule
\caption{The exact rank-two weighted Kemeny algorithm.}
\Description{Pseudocode that closes the two-dimensional factors, sorts them by
polar angle, rotates after the closing vector, and returns the item order.}
\label{alg:closesort}
\end{algorithm}

The polar comparator separates the two half-planes and then orders vectors
by the sign of their determinant; it does not require trigonometric evaluation.
Zero vectors may occupy any position, and same-ray ties may be broken
arbitrarily, for example by radius and identifier.

The area formula and angular edge ordering are
classical~\cite{bader2020horofunction}; we state the precise form for the
Kemeny reduction.

\begin{lemma}[Maximum-area edge ordering]\label{lem:edgeorder}
Let $g_0,\ldots,g_n\in\R^2$ be a multiset with $\sum_tg_t=0$.  Among its cyclic
orders, the doubled signed area of the closed polygon traversed by these edge
vectors is maximized when the vectors occur in cyclic polar-angle order.
\end{lemma}

\begin{proof}
For any cyclic order, set $x_0=0$ and $x_{t+1}=x_t+g_t$.  The vectors close, and
the polygon's doubled signed area is
\begin{equation}
  2A=\sum_t\detv{x_t}{x_{t+1}}
     =\sum_{a<b}\detv{g_a}{g_b}.
 \label{eq:area}
\end{equation}
Cyclic rotations leave~\eqref{eq:area} unchanged because $\sum_tg_t=0$.
There are finitely many cyclic orders, so a maximizing order exists.
If the multiset has at most two vectors, every order has zero area and the
claim is immediate; hence assume that it has at least three vectors.

First suppose the vectors are in general position: whenever two nonempty
consecutive edge blocks have a nonempty complementary block, their vector sums
have nonzero determinant.  Take two consecutive blocks in a maximizing
order, with sums $a,b$.  Swapping them changes~\eqref{eq:area} by
$\det(b,a)-\det(a,b)=-2\det(a,b)$, so $\det(a,b)\ge0$ and, under the
general-position assumption, $\det(a,b)>0$.  Now take any three distinct
vertices $x_p,x_q,x_r$ in cyclic traversal order.  The chords $x_q-x_p$ and
$x_r-x_q$ are sums of two consecutive nonempty blocks, with the remaining
edges forming their nonempty complement.  Hence
\[
  \det(x_q-x_p,x_r-x_q)>0.
\]
In particular, every other vertex lies strictly to the left of each oriented
polygon edge.  Thus the vertices occur counterclockwise on the boundary of
their convex hull: the polygon is simple and strictly convex, and its edge
directions occur in cyclic polar-angle order.

For an arbitrary multiset, take general-position zero-sum perturbations
$g_t^{(k)}\to g_t$.  They exist because the excluded determinant equalities
form a finite union of proper algebraic subsets of the zero-sum subspace.  For
every $k$, the preceding argument makes a polar order optimal.  Along a
subsequence, the same labeled cyclic order is chosen; taking limits
in~\eqref{eq:area} shows that this limiting polar order is optimal for the
original vectors.  Any two polar orders of the original multiset differ only
by moving zero vectors or reordering same-ray vectors.  Such moves have zero
determinant and leave~\eqref{eq:area} unchanged.  Hence every cyclic polar order
is optimal, including all degenerate cases.
\end{proof}

\begin{theorem}[Exact weighted aggregation]\label{thm:exact}
If $W$ is skew-symmetric with $\operatorname{rank}(W)\le2$, \CloseSort
returns a global maximizer of $\J_W$ and hence of~\eqref{eq:likelihood}.  It runs
in $O(n\log n)$ time from factors and $O(n^2+n\log n)$ time from a dense $W$.
\end{theorem}

\begin{proof}
For an item order $\pi$, rotate the augmented edge cycle to
$(h_0,h_{\pi_1},\ldots,h_{\pi_n})$.  Because $h_0=-\sum_i h_i$, the total
contribution of terms containing $h_0$ in~\eqref{eq:area} is zero.  The remaining
terms are exactly $\J_W(\pi)$.  Lemma~\ref{lem:edgeorder} therefore makes the augmented angular
cycle globally optimal, and cutting it after the distinguished edge $h_0$
returns the corresponding item order.  Computing $h_0$ is linear and polar
sorting is $O(n\log n)$.  For a dense exact rank-two $W$,
Proposition~\ref{prop:factors} adds $O(n^2)$ factor recovery and rank
verification.
\end{proof}

\begin{corollary}[Bradley--Terry sorting]\label{cor:bt}
For $W_{ij}=s_i-s_j$, \CloseSort returns the nonincreasing order of the
scores $s_i$.
\end{corollary}
\begin{proof}
Take $h_i=(s_i,1)$ in~\eqref{eq:detfactor}.  All item vectors lie in the open
upper half-plane, and
$\det(h_i,h_j)=s_i-s_j$, so their counterclockwise polar order is precisely
nonincreasing score order.  The closing vector
$h_0=(-\sum_i s_i,-n)$ lies in the lower half-plane; cutting the angular cycle
after $h_0$ therefore starts with the largest score.  Theorem~\ref{thm:exact}
then gives the claim, with arbitrary order among tied scores.
\end{proof}
Rank two strictly extends Bradley--Terry: the factors $h_i$ are no longer
confined to an affine line, and the induced preferences can contain directed
cycles.

\section{Exact gap and residual stability}
\label{sec:gap}

Let $W_0$ be skew rank two, let $\pi_0$ be its \CloseSort order, and
rotate the augmented angular cycle so that
\[
  (g_0,g_1,\ldots,g_n)
  =(h_0,h_{\pi_{0,1}},\ldots,h_{\pi_{0,n}}).
\]
Let $\delta$ be the smallest determinant of two consecutive edges, and assume
that it is positive:
\begin{equation}
 \delta=\min_{0\le t\le n}\detv{g_t}{g_{t+1}}>0,
 \quad g_{n+1}=g_0.
 \label{eq:delta}
\end{equation}
This condition makes the optimal polygon strictly convex and the optimum
unique.

The gap and perturbation results have a common scale.  Geometrically, $\delta$
is the cheapest local way to break the optimal angular cycle: swapping the two
consecutive edges with the smallest determinant loses $2\delta$ in $\J$ and
$\delta$ in log-likelihood.  An arbitrary residual can recover at most twice
its upper-triangle $\ell_1$ norm in $\J$, which gives the regret bound below;
comparing this norm with $\delta$ gives the stability certificate.

\begin{lemma}[Minimum-triangle lemma~\cite{toussaint1982simple}]
\label{lem:mintriangle}
Among the triangles spanned by vertices of a convex polygon, a minimum-area
triangle has two sides that are polygon edges; equivalently, it is an ear formed
by three consecutive vertices.
\end{lemma}

\begin{theorem}[Runner-up gap and residual stability]\label{thm:stability}
Under~\eqref{eq:delta}, let $\pi_0^{(2)}$ have the best $W_0$ score strictly
below $\pi_0$.  Then
\begin{equation}
 \J_{W_0}(\pi_0)-\J_{W_0}(\pi_0^{(2)})=2\delta,
 \quad
 \log\frac{\LL_{W_0}(\pi_0)}{\LL_{W_0}(\pi_0^{(2)})}=\delta.
 \label{eq:gap}
\end{equation}
The second-best score is attained by one of $n-1$ adjacent item transpositions
or the one-step left/right cyclic shift of $\pi_0$.

For $W=W_0+E$ with arbitrary skew residual $E$, write
$\norm{E}_{1,\triangle}=\sum_{i<j}|E_{ij}|$ and let $\pi_W^\star$ optimize $W$.
Then
\begin{equation}
 \log\LL_W(\pi_W^\star)-\log\LL_W(\pi_0)
 \le \norm{E}_{1,\triangle}.
 \label{eq:regret}
\end{equation}
If $\norm{E}_{1,\triangle}<\delta$, then $\pi_0$ is the unique global optimum
for $W$.  The threshold $\delta$ is worst-case tight over skew residuals.
\end{theorem}

\begin{proof}
\emph{Runner-up gap.}
Any cyclic edge order other than the counterclockwise angular order fails the
left-half-plane characterization used in the proof of
Lemma~\ref{lem:edgeorder}; otherwise
it would be the same weakly convex angular cycle and hence optimal.  Thus some
ordered vertex triple is clockwise.  The two corresponding consecutive edge
blocks have sums $a,b$ with $\det(a,b)<0$, so swapping them strictly improves
the score.
If the starting order has the second-best score, this swap must reach the
unique optimum; otherwise it creates a score strictly between second and
first.

Reading the argument in reverse, a runner-up is obtained by swapping two
consecutive edge blocks with sums $a,b$, losing $2\det(a,b)$.  If the blocks run
between consecutive selected vertices $x_p,x_q,x_r$ of the optimal polygon,
then $a=x_q-x_p$, $b=x_r-x_q$, and $\det(a,b)$ is twice the area of triangle
$(x_p,x_q,x_r)$.  Lemma~\ref{lem:mintriangle} lower-bounds it by the minimum
ear determinant $\delta$.  Swapping the two edges of a minimum ear attains that
bound.  An interior ear swaps two adjacent items; the two ears incident to
$h_0$ are the left and right cyclic shifts.
Equation~\eqref{eq:likelihood-reduction} converts the $\J$ gap to the
log-likelihood gap.

\smallskip\noindent\emph{Regret bound.}
Changing an order reverses a subset of unordered pairs.  Each reversed residual
entry changes $\J_E$ by at most $2|E_{ij}|$.  Since $\pi_0$ maximizes
$\J_{W_0}$,
$\J_W(\pi_W^\star)-\J_W(\pi_0)\le2\norm{E}_{1,\triangle}$, which gives
\eqref{eq:regret}.  Every nonoptimal order loses at least $2\delta$ under $W_0$
by the first part; the residual can recover strictly less than that if
$\norm{E}_{1,\triangle}<\delta$.

\smallskip\noindent\emph{Tightness.}
Take the minimum-ear neighbor, which loses $2\delta$
in $\J$, and place a residual of magnitude $\delta$ on any pair it reverses,
opposing $\pi_0$.  The two orders tie; any larger residual flips them.
\end{proof}

Cauchy--Schwarz and skew symmetry give the convenient sufficient condition
\begin{equation}
 \norm{E}_{1,\triangle}
 \le \frac{\sqrt{n(n-1)}}{2}\norm{E}_F<\delta.
 \label{eq:frobstability}
\end{equation}
Writing $\pi=\pi_0$, the gap itself can be computed without geometry:
\begin{equation}
 \delta=\min\left\{\sum_j(W_0)_{\pi_1j},\;
 (W_0)_{\pi_1\pi_2},\ldots,(W_0)_{\pi_{n-1}\pi_n},\;
 -\sum_j(W_0)_{\pi_nj}\right\}.
 \label{eq:directdelta}
\end{equation}
Each interior term $(W_0)_{\pi_k\pi_{k+1}}$ is the half-$\J$ loss from
swapping adjacent items $\pi_k$ and $\pi_{k+1}$; the two boundary terms
correspond to moving the first or last item to the opposite end of the
ranking.

The best Frobenius skew-rank-two approximation follows from the leading
two-dimensional singular subspace~\cite{eckart1936approximation}.  In
particular, write $W_0=\lambda(uv^\top-vu^\top)$ for orthonormal $u,v$ in the
leading invariant subspace and set $h_i=\sqrt{\lambda}(u_i,v_i)$.  Then
$(W_0)_{ij}=\det(h_i,h_j)$.  The regret bound applies to any chosen rank-two
$W_0$; the runner-up identity and stability certificate additionally
require~\eqref{eq:delta}.

\subsection{Exact inference for determinant logits}
\label{sec:detlogit}

Let $\operatorname{sigmoid}(x)=(1+e^{-x})^{-1}$.  Suppose a pairwise model
assigns $h_i\in\R^2$ to every candidate and sets
\begin{equation}
 p_{ij}=\operatorname{sigmoid}(\detv{h_i}{h_j}).
 \label{eq:cyclicmodel}
\end{equation}
We call~\eqref{eq:cyclicmodel} the \emph{determinant-logit} model.  It
coincides with the one-dimensional, bias-free blade-chest-inner model of Chen
and Joachims~\cite{chen2016intransitivity}.  Our result gives the exact
maximum-likelihood list for this model.  Factors may be estimated by binary
cross-entropy from observed comparisons; once supplied,
Algorithm~\ref{alg:closesort} performs exact combinatorial inference.  The
class permits directed cycles and contains scalar Bradley--Terry by
Corollary~\ref{cor:bt}.

This separates \emph{statistical fit} from \emph{combinatorial inference}: for
a rank-two model, inference is exact regardless of how the factors were
estimated.

\section{Separation from sign-only aggregation}
\label{sec:separation}

Prior rank-two MFAST algorithms receive only the tournament signs and minimize
the number of backward arcs~\cite{veerathu2021structure,rajkumar2022theory}.
That objective can oppose maximum likelihood.  Consider factors
\[
h_A=(4,-3),\quad h_B=(1,3),\quad h_C=(-3,-4),\quad h_D=(0,2),
\]
which give
\begin{equation}
W=\begin{pmatrix}
0&15&-25&8\\-15&0&5&2\\25&-5&0&-6\\-8&-2&6&0
\end{pmatrix}.
\label{eq:separation}
\end{equation}
Sign-only MFAST uniquely returns $(A,B,D,C)$, agreeing with five of six edges,
but it has $\J=11$.  \CloseSort and weighted maximum likelihood return
$(C,A,B,D)$, agreeing with only four signs but achieving $\J=39$.

For each $n$ in Table~\ref{tab:separate}, we draw 2,000 independent Gaussian
factor sets.  The sign-only method is solved exactly using the
circular-order/cyclic-shift characterization from prior
work~\cite{rajkumar2022theory}; the weighted method uses \CloseSort.
The sign-optimal order is suboptimal for the weighted objective on 14.7\% of
four-item instances and 63.6\% at $n=32$.  Thus the distinction between
sign-only and weighted optimization becomes more frequent as the number of
items grows.

\begin{table}[t]
\centering
\caption{Frequency with which the sign-optimal order is suboptimal for the
weighted objective, over 2,000 random rank-two instances per size.}
\label{tab:separate}
\small
\begin{tabular}{rrrrrrr}
\toprule
$n$ & 4 & 6 & 8 & 10 & 16 & 32\\
\midrule
Suboptimal (\%) & 14.7 & 22.9 & 28.6 & 35.5 & 46.0 & 63.6\\
\bottomrule
\end{tabular}
\end{table}

\section{Experiments}
\label{sec:experiments}

The experiments address four questions.
\begin{enumerate}[leftmargin=*,itemsep=1pt,label=\textbf{(RQ\arabic*)}]
\item Do exhaustive enumeration and exact optimization match the predicted
  optimum, runner-up gap, and stability threshold?
\item When the population log-odds are rank two, does rank-two projection
  improve finite-sample ranking over plug-in and score-based alternatives?
\item How does \CloseSort perform on real LLM-judge profiles, including those
  with directed cycles?
\item What is the runtime advantage of \CloseSort over exact dynamic programming
  when factors are supplied?
\end{enumerate}

\subsection{Experimental setup}

\textbf{Exactness verification.}  We draw Gaussian factors for
$n=3,\ldots,9$ and
enumerate every order.  Across 419 instances, we compare both the optimum and
the runner-up score with Theorems~\ref{thm:exact} and~\ref{thm:stability}.

\textbf{Residual perturbations.}  We draw Gaussian rank-two factors, add skew
Gaussian residuals whose upper-triangle $\ell_1$ norm ranges from $0.1\delta$ to
$64\delta$, and recover the global optimum with an exact MILP solver.  We
use 25 trials for every residual level and $n\in\{8,12,16\}$.  Targeted residuals
supported on a pair reversed by the runner-up order test the sharp
threshold.

\textbf{Well-specified finite-sample estimation.}  Across five independent
seeds, we generate 1,024 latent $n=14$ profiles per seed (5,120 total).  For
each profile, we sample two-dimensional Gaussian factors at a fixed signal
scale, defining population log-odds $W^\star_{ij}=\det(h_i,h_j)$ and probabilities
$p_{ij}=\operatorname{sigmoid}(W^\star_{ij})$.  For each profile and
$m\in\{4,16,64\}$, we independently sample $m$ Bernoulli comparisons per pair.
With $X_{ij}$ wins by $i$, we form the add-half empirical log-odds
$\widehat W_{ij}=\log\frac{X_{ij}+1/2}{m-X_{ij}+1/2}$, corresponding to
Jeffreys smoothing~\cite{jeffreys1946invariant}, and fit their best
skew-rank-two approximation.  We compare its \CloseSort order with Borda,
Copeland, and the empirical plug-in optimum, computed by exact DP on
$\widehat W$.  Every order is evaluated by regret on
$W^\star$, thereby measuring recovery of the population objective.

\textbf{LLM-judge profiles.} We use Arena-Hard v0.1~\cite{li2025arena} and its
TokenProbs scores released by JuStRank~\cite{gera2025justrank}, covering 63
systems and ten LLM judges.  We
use 100 questions to construct fixed ability-matched cohorts and a disjoint set
of 200 questions for evaluation.  We aggregate the construction split over
judges, sort systems by Borda row sum, and partition adjacent systems into nine
nonoverlapping cohorts of seven.  The evaluation split is partitioned into 40
five-question blocks.  Crossing
blocks with judges and cohorts gives $40\times10\times9=3{,}600$ profiles, each
containing one fixed judge and seven similarly performing systems.  The full
cohort assignment is reported in Table~\ref{tab:cohorts}.

Within a profile, each question contributes a win, loss, or half-tie for a pair
according to its two TokenProbs quality scores.  If $w_{ij}$ and $w_{ji}$ are
the resulting counts, we set
$W_{ij}=\log((w_{ij}+1/2)/(w_{ji}+1/2))$.

Each profile defines a complete empirical log-odds matrix and its weighted
Kemeny objective.  Held--Karp DP provides the exact optimum; the methods below
differ in whether they order the full matrix or its rank-two representation.

\textbf{Baselines and metrics.} For an empirical matrix $W$, $W_2$ denotes the
best Frobenius skew-rank-two approximation, and \CloseSort uses its
two-dimensional factors.  Captured energy is
$E_2=1-\norm{W-W_2}_F^2/\norm{W}_F^2$.  On LLM-judge profiles, the
full-data baselines are Bradley--Terry, Borda, Copeland, Rank Centrality, and
Schulze~\cite{bradley1952rank,zwicker2016voting,negahban2017rank,schulze2011new}.
Borda ranks by row sums of
$W$, and Copeland by row sums of $\operatorname{sign}(W)$.  Factor-Borda and
factor-Copeland apply these score rules to $W_2$, while the sign-only circular
cut selects the best cyclic shift of the factor angular order under
$\operatorname{sign}(W_2)$.

Let $s(W)=\sum_{i<j}|W_{ij}|$, and let $\pi^\star$ be the exact maximizer of
$\J_W$.  The normalized regret of an order $\pi$ is
\[
r_W(\pi)=\frac{\J_W(\pi^\star)-\J_W(\pi)}{s(W)},
\]
and the normalized gain over baseline $b$ is
\[
g_b=\frac{\J_W(\pi_{\rm CS})-\J_W(\pi_b)}{s(W)}
    =r_W(\pi_b)-r_W(\pi_{\rm CS}).
\]
Comparisons on $W_2$ use $s(W_2)$ analogously.  We call a profile
empirical-cyclic when the directed majority graph induced by $W$ contains a
directed triangle, and empirical-acyclic otherwise.  Confidence bounds use 10,000
bootstrap samples of the 40 question blocks, preserving the judges and cohorts
sharing each block, with Bonferroni correction across all comparisons in
Table~\ref{tab:llmjudge}.

\textbf{Runtime benchmark.} We measure ranking-stage inference from supplied
two-dimensional factors.  For $n=6,8,\ldots,18$, we draw 25 independent
factor sets with $h_i\sim\mathcal{N}(0,I_2)$, defining
$W_{ij}=\det(h_i,h_j)$.  Both methods start from the same factors:
\CloseSort times closing-vector construction, angular sorting, and
order reconstruction, while the exact baseline times $W$ materialization,
Held--Karp DP, and reconstruction.  After an untimed warm-up at each $n$, we
collect five single-threaded repeats per instance, timing \CloseSort
in batches of 256 and DP individually.  Reported times first take the median
over repeats and then over the 25 instances; speedups are medians of paired
per-instance ratios.

\subsection{Exactness and residual stability}

\begin{figure}[t]
  \centering
  \includegraphics[width=\linewidth]{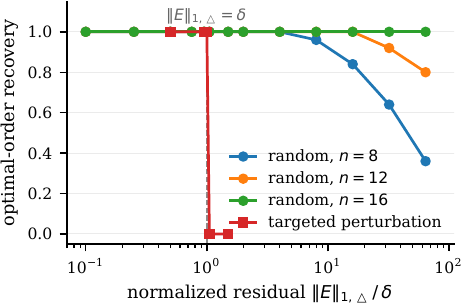}
  \caption{Fraction of trials in which the rank-two \CloseSort order remains
  globally optimal under random or targeted residuals.  The targeted
  construction attains the theorem's threshold.}
  \Description{Recovery curves for CloseSort under random perturbations and a
  targeted worst-case perturbation for several problem sizes.}
  \label{fig:recovery}
\end{figure}

All 419 enumerated optima and runner-up gaps agree with the formulas to
$10^{-7}$.  Figure~\ref{fig:recovery} reports recovery under diffuse random
residuals.  Every tested instance up to $4\delta$ retains the rank-two optimum.
At $64\delta$, recovery is 36\%, 80\%, and 100\% for $n=8,12,16$; distributing
fixed $\ell_1$ mass over more pairs makes the residual less concentrated.
Targeted perturbations produce a tie at $\delta$ and change the optimum at
$1.05\delta$, attaining the threshold in Theorem~\ref{thm:stability}.

\subsection{Finite-sample estimation under the determinant-logit model}

The latent majority tournament is cyclic in 99.88\% of the 5,120 profiles.
Table~\ref{tab:generative} reports mean population regret and \CloseSort's
win rate against plug-in across all five seeds.
Rank-two projection regularizes the empirical estimate toward the latent
determinant-logit structure and has lower population regret in 68.6\%, 84.7\%,
and 85.7\% of profiles at $m=4,16,64$, respectively.  As $m$ grows, the
empirical matrix moves closer to rank two ($E_2$ rises from 47.6\% to 87.7\%),
and both the win rate and the relative reduction in mean regret increase.

\begin{table}[t]
\centering
\caption{Finite-sample study over 5,120 latent determinant-logit profiles
($n=14$) with $m$ observations per pair.  $E_2$: log-odds energy captured by
rank two.  Plug-in: exact optimum of the empirical log-odds.  Win vs.\ plug-in is
the percentage of profiles on which \CloseSort has lower population regret.}
\label{tab:generative}
\small
\setlength{\tabcolsep}{3.5pt}
\begin{tabular}{rr rrrr r}
\toprule
& & \multicolumn{4}{c}{Mean population regret} & \\
\cmidrule(lr){3-6}
$m$ & $E_2$ (\%) & CloseSort & Plug-in & Borda & Copeland &
\shortstack{Win vs.\\plug-in (\%)}\\
\midrule
4  & 47.6 & \textbf{11.509} & 13.993 & 20.812 & 21.003 & 68.6\\
16 & 68.1 & \textbf{3.021}  & 5.847  & 17.506 & 18.467 & 84.7\\
64 & 87.7 & \textbf{.748}   & 1.891  & 15.121 & 16.476 & 85.7\\
\bottomrule
\end{tabular}
\end{table}

\subsection{LLM-judge aggregation on ability-matched cohorts}
\label{sec:llmjudge}

Directed cycles are common in the empirical LLM-judge profiles.  The directed
majority graph induced by $W$ contains a directed triangle in 1,249 of the 3,600
profiles (34.69\%).  The fraction ranges from 22.25\% to 40.75\% across the
nine ability-ordered cohorts.  The rank-two projection $W_2$ captures 89.16\%
of the squared
Frobenius energy overall and 84.18\% within the empirical-cyclic profiles, and
it remains cyclic in 103 of those 1,249 profiles.

Table~\ref{tab:llmjudge} separates the effects of the representation and the
ordering rule.  The first block compares \CloseSort with the full-data
baselines, evaluating every order on
$\J_W$.  The second block holds the factors of $W_2$ fixed and again evaluates
on $\J_W$, isolating the ordering rule.  The final block evaluates the same
factor-based orders on $\J_{W_2}$.

\begin{table}[t]
\centering
\caption{Mean per-profile normalized objective gain of \CloseSort
over each baseline.  \CloseSort uses the factors of $W_2$ throughout;
block headings specify the baseline input and evaluation objective.  Cycle
status is determined by the directed majority graph of the empirical matrix $W$
(1,249 cyclic and 2,351 acyclic of 3,600 profiles).
Positive values favor \CloseSort.}
\label{tab:llmjudge}
\small
\setlength{\tabcolsep}{9pt}
\begin{tabular}{lrrr}
\toprule
Baseline & All & \multicolumn{2}{c}{Empirical $W$}\\
\cmidrule(lr){3-4}
& & Cyclic & Acyclic\\
\midrule
\multicolumn{4}{l}{\itshape Full-data baselines; evaluated on $W$}\\
Bradley--Terry & $.01959$ & $.03307$ & $.01243$\\
Borda & $.02803$ & $.04328$ & $.01993$\\
Copeland & $-.00102$ & $.01567$ & $-.00989$\\
Rank Centrality & $.02310$ & $.03660$ & $.01593$\\
Schulze & $.00320$ & $.04122$ & $-.01700$\\
\addlinespace
\multicolumn{4}{l}{\itshape Same-factor baselines; evaluated on $W$}\\
Factor-Borda & $.03198$ & $.05039$ & $.02220$\\
Factor-Copeland & $.00486$ & $.01142$ & $.00138$\\
Sign-only cut & $.00022$ & $.00107$ & $-.00023$\\
\addlinespace
\multicolumn{4}{l}{\itshape Same-factor baselines; evaluated on $W_2$}\\
Factor-Borda & $.03441$ & $.05669$ & $.02257$\\
Factor-Copeland & $.00524$ & $.01289$ & $.00118$\\
Sign-only cut & $.00030$ & $.00088$ & $.00000$\\
\bottomrule
\end{tabular}
\end{table}

On the 1,249 empirical-cyclic profiles, \CloseSort has mean normalized
regret 0.05531 to the exact empirical optimum, compared with 0.09859 for
Borda.  This is an absolute normalized gain of 0.04328, corresponding to a
43.9\% reduction in regret.  Against the five full-data baselines, the gains
range from 0.01567 over Copeland to 0.04328 over Borda.  All eight
empirical-objective gains in the first two table blocks are positive, and their
Bonferroni-adjusted one-sided block-bootstrap lower bounds are all above zero.

The same-factor comparisons isolate the contribution of the ordering rule.
On empirical-cyclic profiles, the gain ranges from 0.00107 over the sign-only
cut to 0.05039 over Factor-Borda.  In particular, the positive gain over the
sign-only cut measures the benefit of retaining determinant magnitudes when
converting the shared rank-two representation into an order.  On $W_2$ itself,
Theorem~\ref{thm:exact} guarantees that
\CloseSort dominates all three factor-based alternatives; the final
table block quantifies their exact losses on the rank-two objective.

The advantage becomes largest when cyclicity is retained by the rank-two
representation.  The projection $W_2$ is itself cyclic in 118 profiles, 103
of which are also empirical-cyclic.  On this subset, the gains over the five
full-data methods rise to 0.08378--0.15143, and the adjusted lower bounds
remain positive for all eight empirical-objective comparisons.

Across all 3,600 profiles, seven of the eight empirical-objective gains are
positive; the exception is full-data Copeland, which exceeds
\CloseSort by 0.00102 on average.  In the empirical-acyclic profiles,
Copeland and Schulze attain higher empirical objectives, while the same-factor
sign-only cut nearly ties \CloseSort; the remaining
empirical-objective comparisons favor \CloseSort.  The advantage is
therefore concentrated in profiles whose observed pairwise relation is cyclic.

\subsection{Runtime}

\begin{figure}[t]
  \centering
  \includegraphics[width=\linewidth]{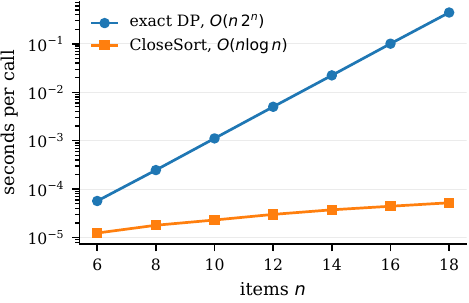}
  \caption{Single-thread ranking-stage runtime from supplied two-dimensional
  factors.  Each point is the median over 25 exact Gaussian rank-two instances
  after taking the median of five repeats within each instance.}
  \Description{Log-scale ranking-stage runtime curves for exact dynamic
  programming and CloseSort, aggregated over repeated measurements.}
  \label{fig:scaling}
\end{figure}

Figure~\ref{fig:scaling} shows runtime scaling from supplied factors.
Exact DP grows from $57\,\mu$s at $n=6$ to $0.442$~s at $n=18$, consistent
with $O(n2^n)$, while \CloseSort grows from $12$ to $52\,\mu$s,
consistent with $O(n\log n)$.  The median paired speedup reaches 2,247-fold at
$n=16$ and 8,407-fold at $n=18$.

\section{Related work}

\textbf{Polygon edge rearrangement.}  For prescribed planar edge vectors, the
signed area is a sum of pairwise determinants and angular ordering maximizes
that area.  Bader and Finkelshtein use both facts, together with the closing
edge $-\sum_i h_i$, in their ordered half-polygon
construction~\cite{bader2020horofunction}.  We use this known geometric result as the
optimization primitive in Theorem~\ref{thm:exact}; the rank-aggregation
reduction, distinguished cut, exact runner-up, and residual analysis are the
steps developed here.

\textbf{Kemeny and weighted tournaments.} Kemeny's rule aggregates rankings by
pairwise disagreement~\cite{kemeny1959mathematics,young1988condorcet}.  Its
weighted form is a linear ordering problem on a weighted tournament; Charon
and Hudry survey the resulting formulations, complexity, and
algorithms~\cite{charon2010updated}.  The exact problem is hard
even under severe input
restrictions~\cite{bartholdi1989voting,charbit2007minimum}.  General
weighted tournaments admit approximation algorithms based on randomized
pivoting, PTASs, and useful score
orderings~\cite{ailon2008aggregating,kenyon2007ptas,coppersmith2010ordering}; exact
Kemeny computation has also been developed through search and admissible
bounds and through fixed-parameter
algorithms~\cite{davenport2004computational,conitzer2006improved,betzler2009fixed}.  Recent work revisits fast combinatorial
solvers~\cite{vahidi2024minimum}.  Our result gives an exact
sorting-based algorithm for an algebraically structured weighted class.

\textbf{Low-rank pairwise models.} Classical paired-comparison models such as
Thurstone and Bradley--Terry derive probabilities from scalar latent strengths
and induce a transitive population
order~\cite{thurstone1927law,bradley1952rank}.  More general statistical formulations
explicitly accommodate systematic
intransitivity~\cite{cattelan2012models,makhijani2019parametric,spearing2023modeling}.
HodgeRank decomposes pairwise data into gradient,
cyclic, and harmonic components~\cite{jiang2011hodge}.  Rajkumar and Agarwal
study sample-efficient ranking under link-low-rank assumptions, while Shah et
al. analyze the statistical and computational consequences of stochastic
transitivity~\cite{rajkumar2016lowrank,shah2016stochastically}.  Causeur and Husson develop a
two-dimensional nontransitive Bradley--Terry
extension~\cite{causeur2005extension}.  Chen and Joachims introduce blade-chest models
for intransitive pairwise prediction; their one-dimensional bias-free
blade-chest-inner logit is
Equation~\eqref{eq:cyclicmodel}~\cite{chen2016intransitivity}.  Our result supplies exact joint-list inference
for this determinant-logit class.

Most directly, Veerathu and Rajkumar and the subsequent tournament-representation
theory characterize sign tournaments induced by rank-two skew matrices and
solve unweighted MFAST with Quicksort and cyclic
shifts~\cite{veerathu2021structure,rajkumar2022theory}.  We retain magnitudes and solve
the likelihood objective; Equation~\eqref{eq:separation} shows that the two
objectives are distinct.  Recent algorithms solve weighted feedback arc set on
the distinct class of cycle-Mengerian tournaments~\cite{song2026weighted}.

\section{Limitations}

The exact reduction applies to complete pairwise objectives: every unordered
pair contributes a determinant term to the polygon-area identity.  With missing
comparisons, this identity no longer represents the observed-edge objective, so
the current construction does not directly extend to sparse observation graphs.
Identifying sparse graph families that preserve an exact geometric reduction
remains open.

The finite-sample experiment assumes a determinant-logit population, making
rank two the correct latent model.  Under misspecification, projection trades
variance reduction against approximation error, and its effect on ranking
consequently depends on the magnitude and structure of the component beyond
rank two.  Characterizing this trade-off beyond the well-specified setting
remains open.

\section{Conclusion}

We introduced \CloseSort, an exact algorithm for
maximum-likelihood aggregation when the skew-symmetric log-odds matrix has rank
two; from the factors, it runs in $O(n\log n)$ time.  By adding a closing
vector and interpreting the rank-two factors as
polygon edges, the same geometry yields the optimal order, identifies the
runner-up gap when $\delta>0$, and gives a tight stability certificate.
This extends the familiar
sort-by-score solution for Bradley--Terry logits to a cyclic determinant-logit
class.

Exact optimization and perturbation experiments match the predicted optimum,
runner-up gap, and stability threshold.  In the finite-sample determinant-logit
setting, rank-two projection lowers population regret relative to the empirical
plug-in optimum, showing that it can improve estimation under sampling noise.
On cyclic LLM-judge profiles, the rank-two
projection captures most of the log-odds energy, while \CloseSort
reduces mean normalized regret by 44\% relative to Borda and yields positive
mean normalized gains against the remaining baselines.  The same rank-two
geometry therefore supports exact inference under the model and remains useful
for aggregation when rank two is only approximate.

\section{Ethical considerations}

This work studies the algorithmic aggregation of pairwise preferences and uses
only publicly released JuStRank scores.  The resulting rankings summarize the
supplied comparisons and may inherit systematic biases or disagreements among
their judges; they should therefore not be interpreted as intrinsic measures of
model quality.

\appendix
\section{Degeneracies}

Theorem~\ref{thm:exact} requires no nondegeneracy assumption.  If $W=0$,
every order is optimal.  More generally, moving a zero vector or permuting
vectors within a same-ray tie in the augmented angular cycle leaves the area
objective unchanged.  Any polar order obtained by a fixed tie-breaking rule is
therefore optimal by Theorem~\ref{thm:exact}.  Antipodal vectors have distinct
polar angles despite their zero mutual determinant; this zero enters
$\delta$ only when the vectors are consecutive in the augmented cycle.

The value of $\delta$ determines which further conclusions hold.  When
$\delta=0$, the angular cycle contains a flat turn, so the optimum need not be
unique and no positive gap to the best alternative is guaranteed.  \CloseSort
remains globally optimal, and the regret bound~\eqref{eq:regret} continues to
hold for an arbitrary skew residual.  The exact runner-up formula and stability
certificate require $\delta>0$.

\section{JuStRank cohorts}

Table~\ref{tab:cohorts} lists the fixed ability-matched cohorts used to
construct the profiles in Section~\ref{sec:llmjudge}.

\begin{table*}[t]
\centering
\caption{Ability-ordered JuStRank cohorts.  Ranks are determined on the
construction split.  Each cohort contributes 400 evaluation profiles from 40
question blocks and ten judges; the final column is the fraction that are
empirically cyclic.}
\label{tab:cohorts}
\small
\setlength{\tabcolsep}{3pt}
\renewcommand{\arraystretch}{1.08}
\begin{tabular}{rr >{\raggedright\arraybackslash\ttfamily\sloppy}p{0.73\textwidth} r}
\toprule
Cohort & Ranks & \multicolumn{1}{c}{Systems} & Cyclic (\%)\\
\midrule
1 & 1--7 & claude-3-5-sonnet-20240620; gpt-4o-2024-05-13; gpt-4-1106-preview; gpt-4-turbo-2024-04-09; mistral-large-2407; yi-large-preview; gpt-4-0125-preview & 29.50\\
2 & 8--14 & claude-3-opus-20240229; gpt-4o-mini-2024-07-18; athene-70b-0725; llama-3.1-405b-instruct-fp8; yi-large; glm-4-0116; claude-3-sonnet-20240229 & 32.50\\
3 & 15--21 & glm-4-0520; gemini-1.5-pro-api-0409-preview; nemotron-4-340b-instruct; gemma-2-27b-it; llama-3.1-405b-instruct-fp8-no-sys-prompt; gemini-1.5-pro-api-0514; llama-3.1-70b-instruct & 34.50\\
4 & 22--28 & gpt-4-0314; mixtral-8x22b-instruct-v0.1; deepseek-coder-v2; mistral-next; llama-3-70b-instruct; mistral-large-2402; qwen2-72b-instruct & 36.75\\
5 & 29--35 & gpt-4-0613; glm-4-air; qwen1.5-72b-chat; claude-3-haiku-20240307; claude-2.0; mistral-medium; claude-2.1 & 36.00\\
6 & 36--42 & command-r-plus; gpt-3.5-turbo-0125; gpt-3.5-turbo-1106; gemini-1.5-flash-api-0514; dbrx-instruct-preview; gpt-3.5-turbo-0613; phi-3-medium-4k-instruct & 40.25\\
7 & 43--49 & phi-3-small-8k-instruct; mixtral-8x7b-instruct-v0.1; snowflake-arctic-instruct; starling-lm-7b-beta; gpt-3.5-turbo-0314; snorkel-mistral-pairrm-dpo; llama-3-8b-instruct & 40.75\\
8 & 50--56 & tulu-2-dpo-70b; yi-34b-chat; llama-3.1-8b-instruct; command-r; mistral-7b-instruct; starling-lm-7b-alpha; gemini-pro & 39.75\\
9 & 57--63 & phi-3-mini-128k-instruct; gemma-1.1-7b-it; vicuna-33b; llama-2-70b-chat; gemma-7b-it; gemma-1.1-2b-it; gemma-2b-it & 22.25\\
\bottomrule
\end{tabular}
\end{table*}

\FloatBarrier

\begin{acks}
The authors used AI tools to assist with manuscript preparation and code
development. All AI-generated content was reviewed and verified by the
authors, who take full responsibility for this work.
\end{acks}

\bibliographystyle{ACM-Reference-Format}
\bibliography{references}

\end{document}